\documentclass{article}
\usepackage{spconf,amsmath,graphicx}

\usepackage{enumitem}
\usepackage{cite}
\usepackage{amsmath,amssymb,amsfonts}
\usepackage{algorithmic}
\usepackage{makecell}
\usepackage{stfloats}
\usepackage{multirow}
\usepackage{subfig}
\usepackage{textcomp}
\usepackage{booktabs}
\usepackage{xcolor}
\setlist{nosep, leftmargin=14pt}

\usepackage{mwe} % to get dummy images

\title{Multi-modal Intermediate Feature Interaction AutoEncoder for Overall Survival Prediction of Esophageal Squamous Cell Cancer}
\name{Chengyu Wu$^{a}$, Yatao Zhang$^{a}$, Yaqi Wang$^{b}$, Qifeng Wang$^{c}$, and Shuai Wang$^{d,e}$}

\address{$^{a}$ Department of Mechanical, Electrical and Information Engineering,\\ Shandong University, Weihai, China \\
    $^{b}$ College of Media Engineering, Communication University of Zhejiang, Hangzhou, China \\
    $^{c}$ Department of Radiation Oncology, Sichuan Cancer Hospital and Institution, Sichuan Cancer Center, \\School of Medicine, Radiation Oncology Key Laboratory of Sichuan Province, \\  University of Electronic Science and Technology of China, Chengdu, China \\
    $^{d}$ School of Cyberspace, Hangzhou Dianzi University, Hangzhou, China\\
    $^{e}$ Suzhou Research Institute of Shandong University, Suzhou, China
    } 

\begin{document}
%\ninept
%
\maketitle
\begin{abstract}
Survival prediction for esophageal squamous cell cancer (ESCC) is crucial for doctors to assess a patient’s condition and tailor treatment plans. The application and development of multi-modal deep learning in this field have attracted attention in recent years. However, the prognostically relevant features between cross-modalities have not been further explored in previous studies, which could hinder the performance of the model. Furthermore, the inherent semantic gap between different modal feature representations is also ignored. In this work, we propose a novel autoencoder-based deep learning model to predict the overall survival of the ESCC. Two novel modules were designed for multi-modal prognosis-related feature reinforcement and modeling ability enhancement. In addition, a novel joint loss was proposed to make the multi-modal feature representations more aligned. Comparison and ablation experiments demonstrated that our model can achieve satisfactory results in terms of discriminative ability, risk stratification, and the effectiveness of the proposed modules.
\end{abstract}
\section{Introduction}
Esophageal cancer (EC) is a prevalent form of cancer that ranks sixth in terms of global mortality rates among all cancer-related cases \cite{sung2021global}. The predominant histological subtype of EC is Esophageal Squamous Cell Cancer (ESCC), which accounts for approximately 90\% of all EC cases \cite{sung2021global}. However, the prognosis for ESCC remains discouraging due to its insidious onset, often resulting in an advanced-stage diagnosis upon initial detection \cite{yang2019ct}. Hence, there is an urgent need to develop an efficient system for predicting the survival situation of ESCC patients to enhance the prognosis.

Deep learning-based image feature extraction methods are widely used and have the ability to automatically extract high-order semantic features from images. In the past few years, lots of autoencoder-based end-to-end deep learning models for survival prediction were proposed \cite{lin2021ct,meng2022deepmts}. Lin et al. \cite{lin2021ct} proposed a novel attention-based framework for evaluating clinical outcomes of esophageal cancer optimized by uncertainty joint loss. However, the above methods utilize only uni-modal data for model construction, which could result in the model's inability to learn complementary information from multiple modalities, hindering both expressive power and robustness. Meng et al. \cite{meng2022deepmts} proposed a multi-task survival model for joint survival prediction and tumor segmentation in advanced nasopharyngeal carcinoma, the features extracted from intermediate layers of the image encoder were concatenated with clinical parameters for further survival prediction usage. However, the processing of the semantic gap between cross-modal features was ignored. Also, the interactive fusion between deep image features and tabular features is not mentioned.

To handle these challenges, we propose a novel framework called Multi-modal Intermediate Feature Interaction AutoEncoder (MIFI-AE) to predict the overall survival of ESCC patients. Two novel modules were devised, named Cross-modal Multi-step Intermediate Fusion Module (CMIFM) for cross-modal feature interactive fusion and Multi-scale Feature map Fusion-Separation Module (MFFSM) for feature decoding reinforcement fusion, respectively. Finally, a novel Multi-task Joint Loss (MJ-Loss) was proposed for model optimization.

\section{Related Works}
For multi-modal prognostic prediction methods, Meng et al. \cite{meng2022deepmts} proposed a novel CT-based multi-modal method for nasopharyngeal carcinoma segmentation and prognosis prediction. A hard-sharing backbone and a cascaded survival network were proposed to extract local features related to the primary tumors. Amini et al. \cite{amini2021multi} developed a multi-modal radiomic model by integrating information extracted from PET and CT images to predict the prognosis of non-small cell lung carcinoma. Feature fusion approaches were applied at feature- and image-levels. Schulz et al. \cite{schulz2021multimodal} developed a multi-modal deep learning model for prognosis prediction in clear-cell renal cell carcinoma. Histopathological, CT, and genomic features were extracted and concatenated for further utilization. 

\section{Method}
\begin{figure*}[htbp]
    \centering
\includegraphics[width=0.84\textwidth]{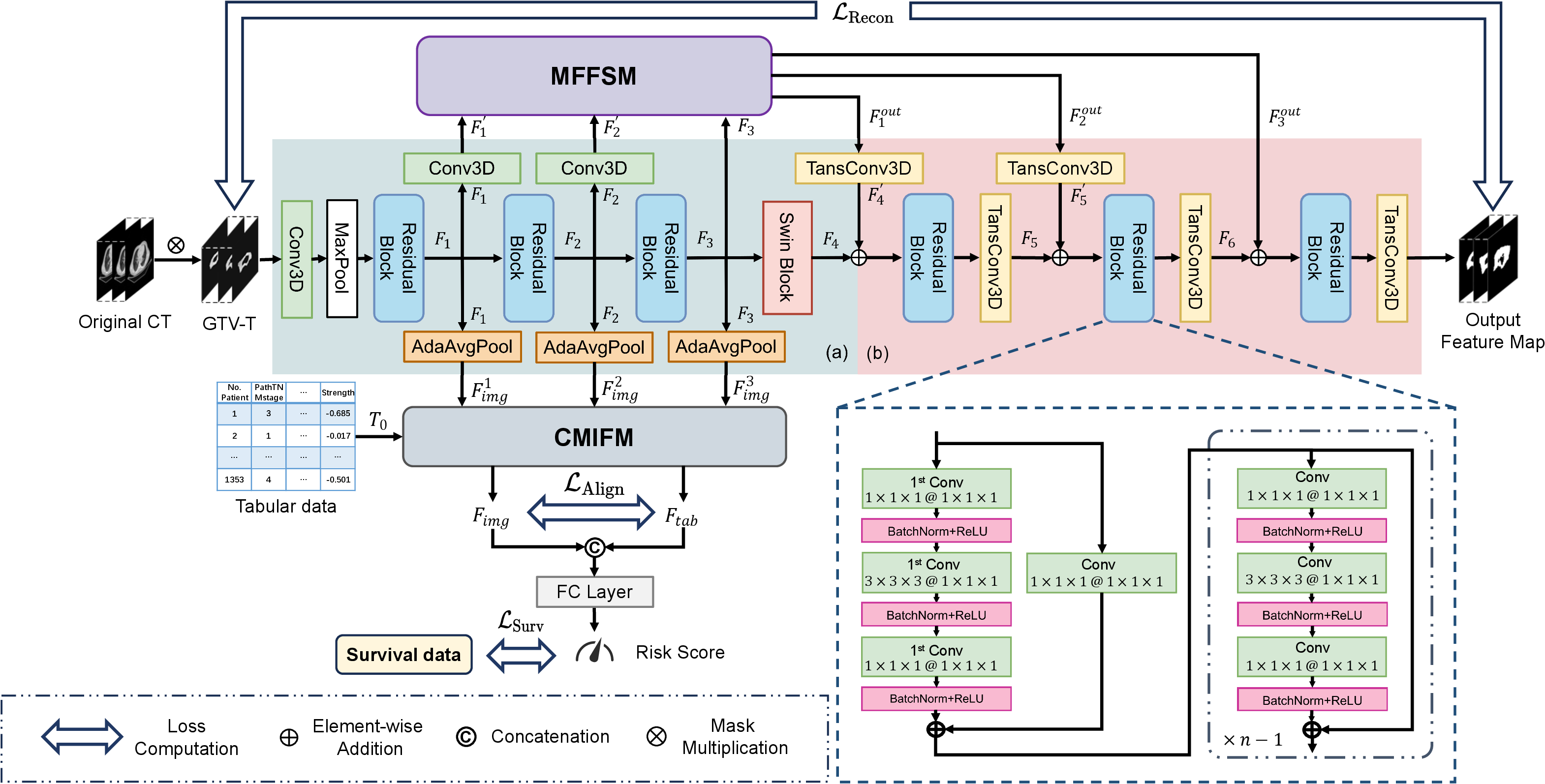}
    \caption{The pipeline of proposed Multi-modal Intermediate Feature Interaction AutoEncoder (MIFI-AE). (a) Encoding part of the MIFI-AE. (b) Decoding part of the MIFI-AE.  }
    \label{MIFI-AE}
\end{figure*}
\subsection{Overview}
The main architecture of the proposed MIFI-AE is shown in Fig. \ref{MIFI-AE}. The encoder and decoder mainly consist of residual blocks stacked by three layers \cite{he2016deep}. At the connection part of the encoder and decoder, a Swin Transformer block was utilized to further enhance the decoding ability of deep feature maps \cite{liu2021swin}. First, the original CT image was multiplied with a tumor mask to obtain the gross target volume of the tumor (GTV-T). Then, the GTV-T was processed by MIFI-AE. The output feature maps of the intermediate layer of the encoder were used for feature decoding reinforcement fusion (through MFFSM) to enhance the decoding ability of the model and also for cross-modal feature interaction (through CMIFM) to generate the final risk score, which is applied for survival prediction.

\subsection{Multi-scale Feature map Fusion-Separation Module (MFFSM)}
\begin{figure}[htbp]
    \centering
\includegraphics[width=0.5\textwidth,keepaspectratio]{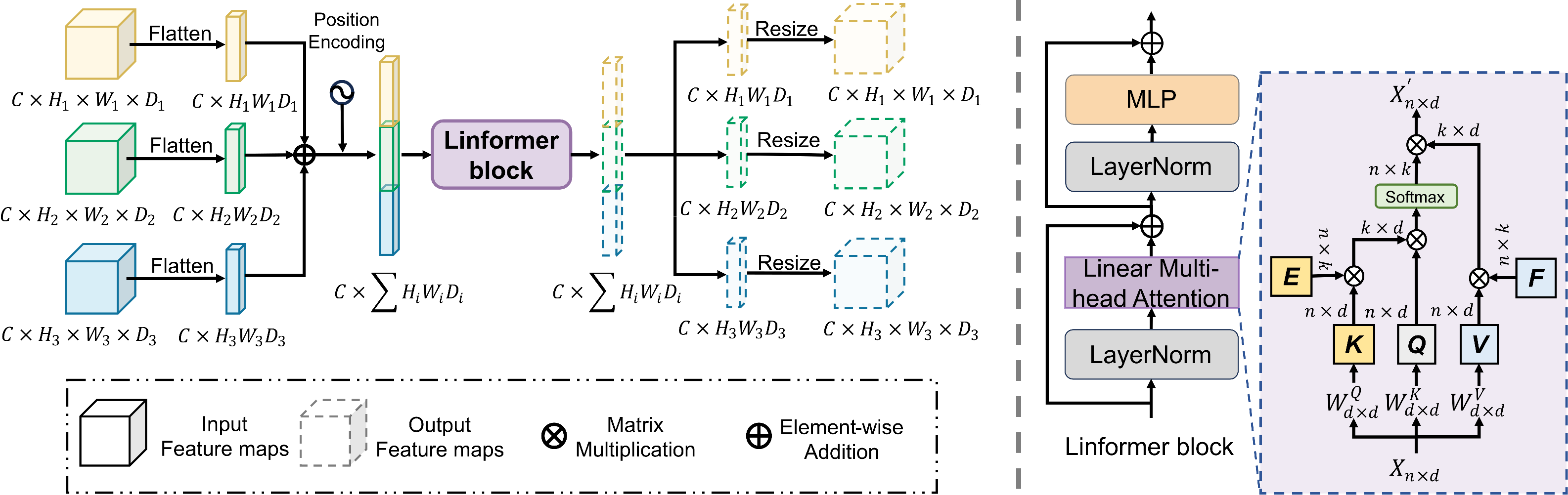}
    \caption{The illustration of proposed Multi-scale Feature map Fusion-Separation Module (MFFSM).}
    \label{MFFSM}
\end{figure}

To better enhance the encoding-decoding ability of the model, we proposed an MFFSM to fuse the information of multi-scale feature maps obtained from intermediate layers of the encoder by flattening and composing them together. After that, the fused feature maps are separated and resized back to their original shape for decoding usage. The whole procedure can be seen in Fig. \ref{MFFSM}. Given input feature maps: $F_{i} \in \mathbb{R}^{C \times H_{i} \times W_{i} \times D_{i}}, \ i \in \{1,2,3\}$. We firstly convert all feature maps into vector form: $F_{i} \in \mathbb{R}^{C \times H_{i} W_{i} D_{i}}, \ i \in \{1,2,3\}$ and concatenate them as $F_{vector} \in \mathbb{R}^{C\times \big( \sum_{i}^{i \in \{1,2,3\}} H_{i} W_{i} D_{i} \big) }$; After that, the position information is encoded into $F_{vector}$. Then, the application of Linear Multi-head Self-Attention \cite{wang2020linformer} is performed on $F_{vector}$. After that, the feature vector can be split into three parts and transformed back to the original input shape of feature maps.

% , which is defined as:
% \begin{align} \label{MHA}
%     &\operatorname{MultiHeadAttention}(Q, K,V) = \notag  \\ &\operatorname{Concat}\left(\operatorname{head}_{1}, \operatorname{head}_{2}, \ldots, \operatorname{head}_{h}\right) W^{O}. 
% \end{align}

% where $Q, K, V \in \mathbb R^{n\times d}$ are input embedding matrices, $n$ is sequence length, $d$ is the embedding dimension, and $h$ is the number of heads. Each head is defined as:
% \begin{align} \label{Attetion}
%     \operatorname { head }_{i} &=\operatorname{LinearAttention}\left(Q W_{i}^{Q}, E_{i} K W_{i}^{K}, F_{i} V W_{i}^{V}\right), \notag \\
%      &=\underbrace{\operatorname{softmax}\left(\frac{Q W_{i}^{Q}\left(E_{i} K W_{i}^{K}\right)^{T}}{\sqrt{d_{k}}}\right)}_{P: \ n \times k} \cdot \underbrace{F_{i} V W_{i}^{V}}_{k \times d}. 
% \end{align}

% where $W^{Q}_{i},W^{K}_{i},W^{V}_{i} \in \mathbb{R}^{d\times d}$ are learned matrices; $E_{i},F_{i} \in \mathbb{R}^{n\times k}$ are projection matrices that project $KW^{K}_{i}$ and $VW^{V}_{i}$ from $\mathbb{R}^{n\times d}$ to $\mathbb{R}^{n \times k}$, we can reduce memory and computation consumption if we choose a much lower k such that $k \ll n$. 

\subsection{Cross-modal Multi-step Intermediate Fusion Module (CMIFM)}

\begin{figure}[htbp]
    \centering
\includegraphics[width=0.78\columnwidth]{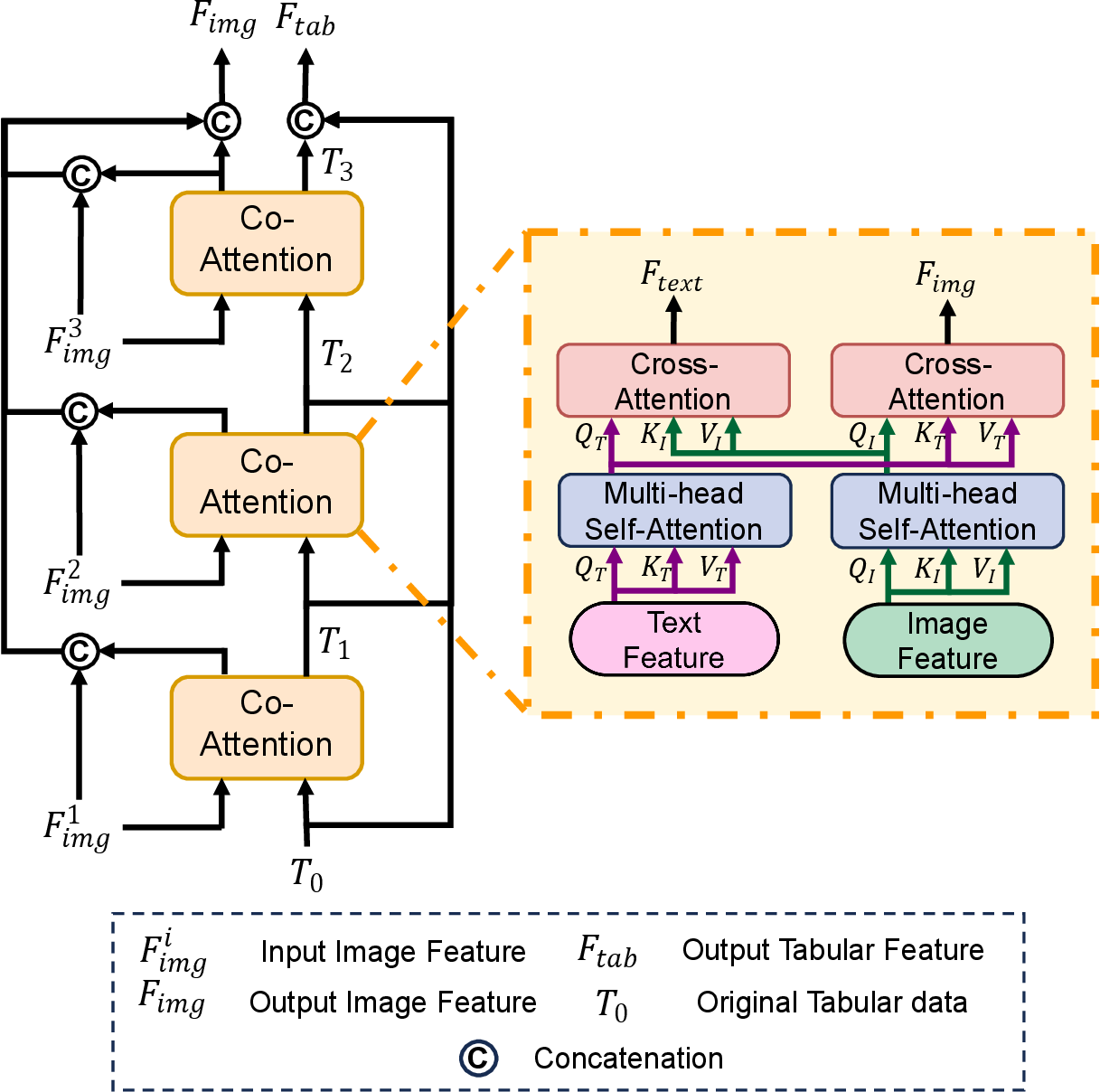}
    \caption{The illustration of proposed Cross-modal Multi-step Intermediate Fusion Module (CMIFM).}
    \label{CMIFM}
\end{figure}

To strengthen the representing capability of prognostic-related features in cross-modal features, the CMIFM was proposed and applied by interactively fusing the cross-modal features with multi-steps repeatedly. The structure of CMIFM is shown in Fig. \ref{CMIFM}, three Co-Attention modules were applied \cite{hendricks2021decoupling} , $F_{img}^{i}, i \in \{1,2,3\}$ stand for the image features, original tabular data $T_{0}$ was first fused with the $F_{img}^{1}$, the output of tabular feature $T_{1}$ was used for next step fusion with $F_{img}^{2}$, the rest may be deduced by analogy. To the end, the output feature of the image and tabular data $F_{img}, F_{tab}$ is further used for loss computation and risk score generation.

\subsection{Multi-task Joint Loss (MJ-Loss)}
For model optimization, we devise a novel MJ-Loss that contains three optimization sub-objectives. For image reconstruction, we applied mean square error loss, which is defined as:
\begin{align}
    \mathcal{L}_{\mathrm{Rec}} = -\frac{1}{N} \sum_{i=1}^{n} \sum_{j=1}^{n}  \sum_{k=1}^{n}\left(I_{T}^{i,j,k}-I_{out}^{i,j,k}\right)^{2}.
\end{align}
where $I_{T}^{i,j,k}$ and $I_{out}^{i,j,k}$ are the voxel values at position $(i,j,k)$ for the input GTV-T and output feature maps from the decoder, respectively. $N$ is donated as the total number of voxels in $I_{T}^{i,j,k}$ and $I_{out}^{i,j,k}$. To eliminate the semantic information gap between different modal representations, we use Kullback-Leibler divergence loss to align image and tabular features, which is defined as:
\begin{align}
    \mathcal{L}_{\mathrm{Align}}(F_{img}\|F_{tab}) = \sum_{i=1}^{M}\Big[f_{img}^{i} \log (f_{img}^{i})-f_{img}^{i} \log (f_{tab}^{i})\Big].
\end{align}
where $f_{img}^{i}$ and $f_{tab}^{i}$ are elements in the final output representations $F_{img}$ and $F_{tab}$ from CMIFM. $M$ donate as the length of $F_{img}$ and $F_{tab}$. For final survival prediction, we derive Cox partial log-likelihood loss, which is defined as:
\begin{align}
    \mathcal{L}_{\mathrm{Surv} }= - \frac{1}{W} \sum_{i: E_{i}=1}\left(h\left(x_{i}\right)-\log \sum_{j: T_{j} \geq T_{i}} e^{h\left(x_{j}\right)}\right).
\end{align}
where $T_{i}$ stands for event time; $E_{i}$ stands for event indicator, where $E_{i} = 1$ means the event was observed. $h(x_{i})$ stands for the risk score of the patient $x_{i}$. $W$ stands for the number of non-censored patients. The final loss can be defined as:
\begin{align}
    \mathcal{L}_{final} =  \mathcal{L}_{\mathrm{Rec}} + \mathcal{L}_{\mathrm{Align}} + \mathcal{L}_{\mathrm{Surv} }.
\end{align}

which is used for final model optimization.

\section{Results}
\subsection{Data Description}
A total of 1,354 ESCC patients were enrolled in this study. Preoperative CT images, clinical hematology parameters, and follow-up information were collected at Sichuan Cancer Hospital. Pathological tumor-node-metastasis (pTNM) stages were determined by The American Joint Committee on Cancer's (AJCC) 8th edition classification system. We conducted a five-fold cross-validation to further test the performance of the model. We randomly selected 80\% of patients as the training cohort (n = 1,083) and the remaining 20\% as the test cohort (n = 271).

\subsection{Performance of Discriminative ability}
For evaluation of models' discriminative ability, we applied the C-index to measure the consistency between predicted risk score and survival status \cite{harrell1982evaluating}. We evaluated MIFI-AE with several other deep learning-based methods. As shown in Table \ref{tab:cindex_os}, our proposed MIFI-AE acquired the best C-index performance of 0.697 $\pm$ 0.01. The second highest performance was obtained from SurvivalCNN, which gets a C-index of 0.693 $\pm$ 0.02.
%As can be seen from Table \ref{tab:cindex_os} models using only uni-modal data performed poorly, while the rest of the models using multi-modal data showed greater robustness on the whole five-fold test set, which illustrates the ability of multi-modal data to provide more complementary prognostically relevant information compared to uni-modal data. 
\begin{table}[htbp]
    \centering
    \caption{C-index comparison of the proposed method and other related methods for OS prediction; $\ast$ stands for p-value $<$ 0.05. $\dagger$ stands for p-value $<$ 0.1.}
     \begin{tabular}{cccc}
  \hline\rule{0pt}{7pt}
   \multirow{2}*{Methods} & \multicolumn{2}{c}{Data included} & \multirow{2}*{\makecell[c]{C-index \\ (Mean $\pm$ Std)}} \\ 
   \cline{2-3}\rule{0pt}{8pt}
  & Image & Tabular   \\
  \hline\rule{0pt}{8pt}
   DeepSurv\cite{katzman2018deepsurv} & & $\checkmark$ & 0.641 $\pm$ 0.01$^\ast$  \\
   DeepMTS\cite{meng2022deepmts}  & $\checkmark$ & $\checkmark$ & 0.690 $\pm$ 0.01 \\
   SurvivalCNN\cite{hao2022survivalcnn}  & $\checkmark$ & $\checkmark$ & 0.693 $\pm$ 0.02  \\
   CACA-UCOM\cite{lin2021ct} & $\checkmark$ &  & 0.608 $\pm$ 0.02$^\ast$  \\
   FullerMDA\cite{naser2021progression}  & $\checkmark$ & $\checkmark$ & 0.678 $\pm$ 0.01$^\dagger$   \\
   
  \hline\rule{0pt}{10pt}
  \textbf{\makecell[c]{MIFI-AE (Ours)}} & $\checkmark$ & $\checkmark$ & \textbf{0.697 $\pm$ 0.02 }  \\
  \hline
    \end{tabular}
    \label{tab:cindex_os}
\end{table}

\subsection{Performance of  Risk Stratification}
\begin{figure}[!ht]
\centering
\subfloat[DeepSurv]{\includegraphics[width=0.165\textwidth]{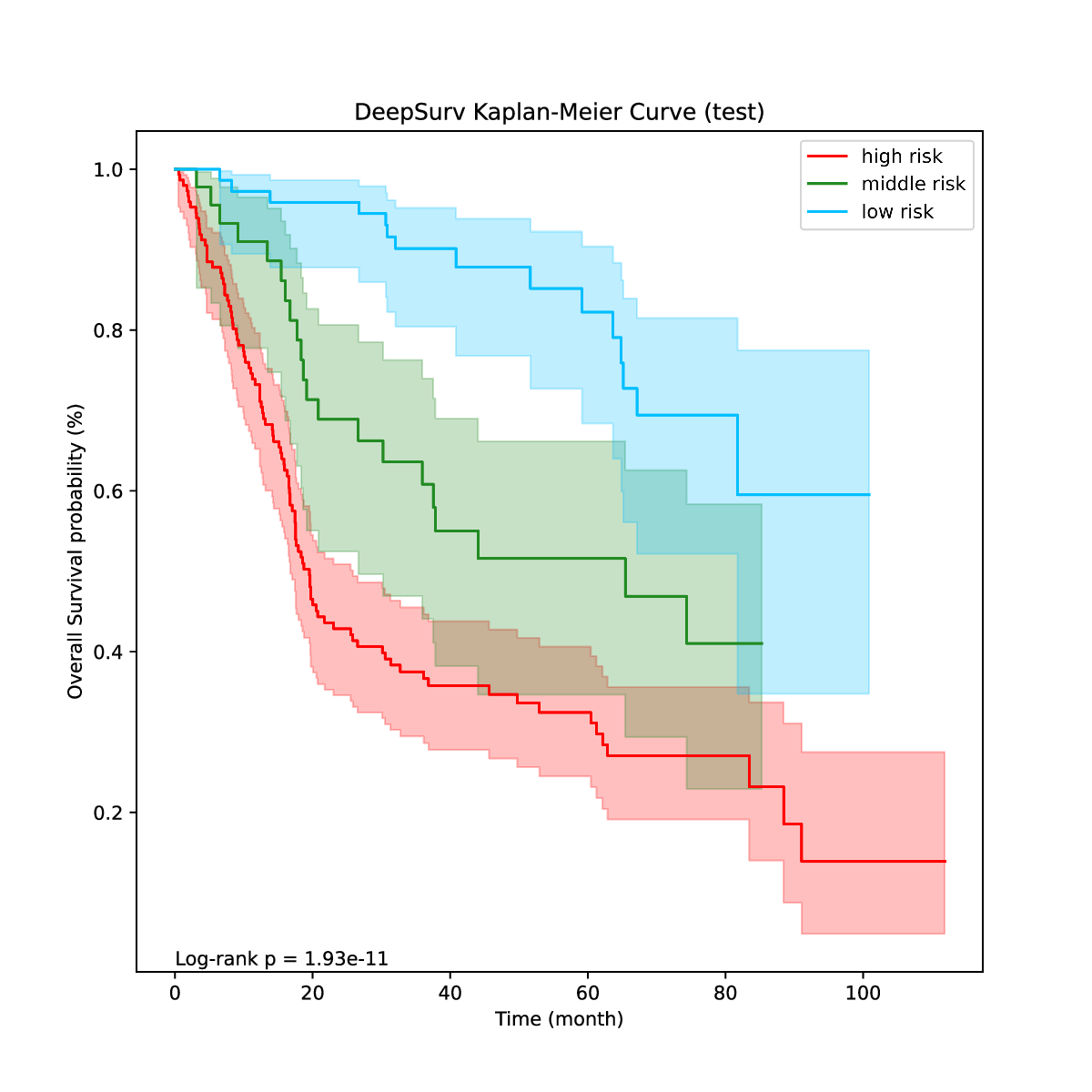}}
\subfloat[DeepMTS]{\includegraphics[width=0.165\textwidth]{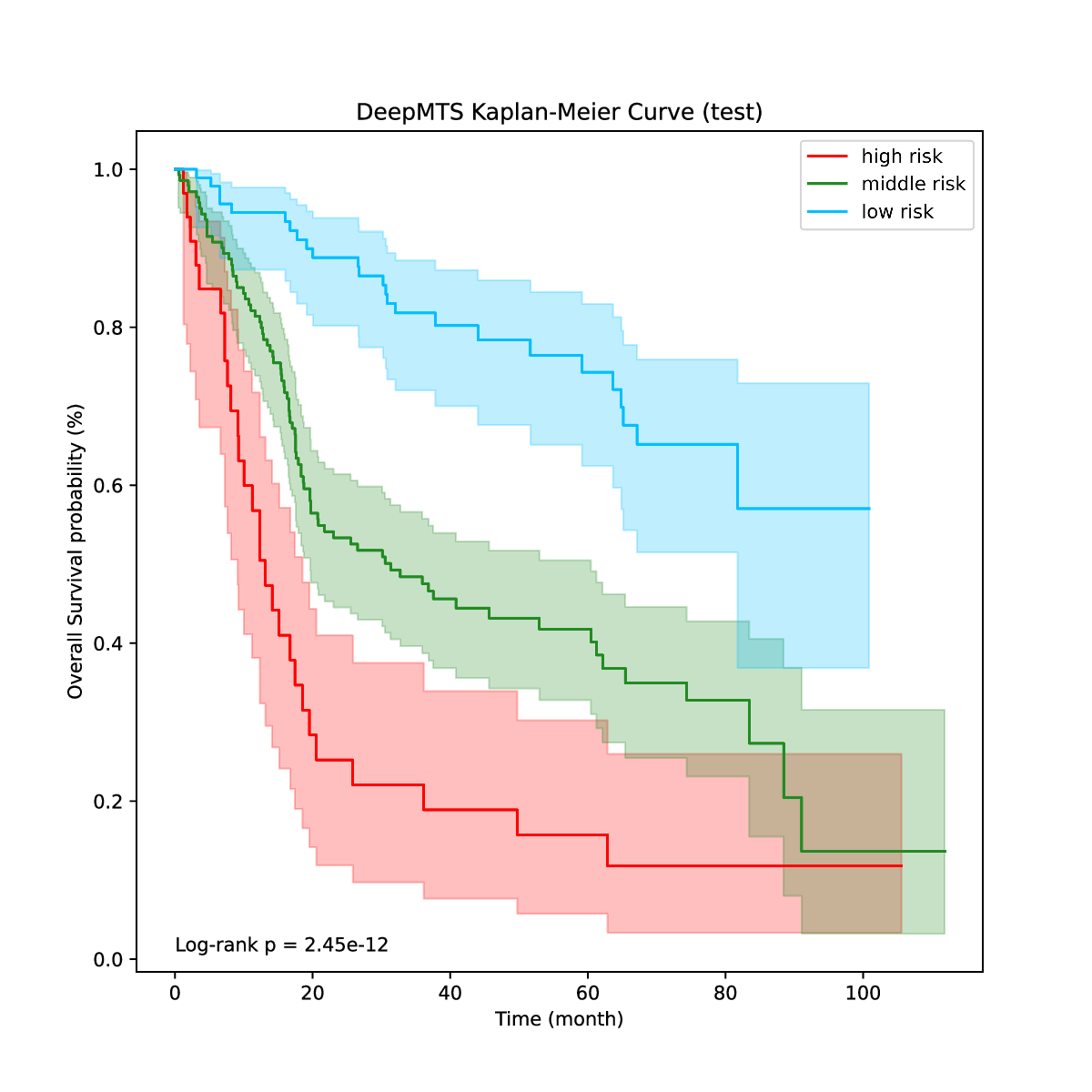}}
\subfloat[SurvivalCNN]{\includegraphics[width=0.165\textwidth]{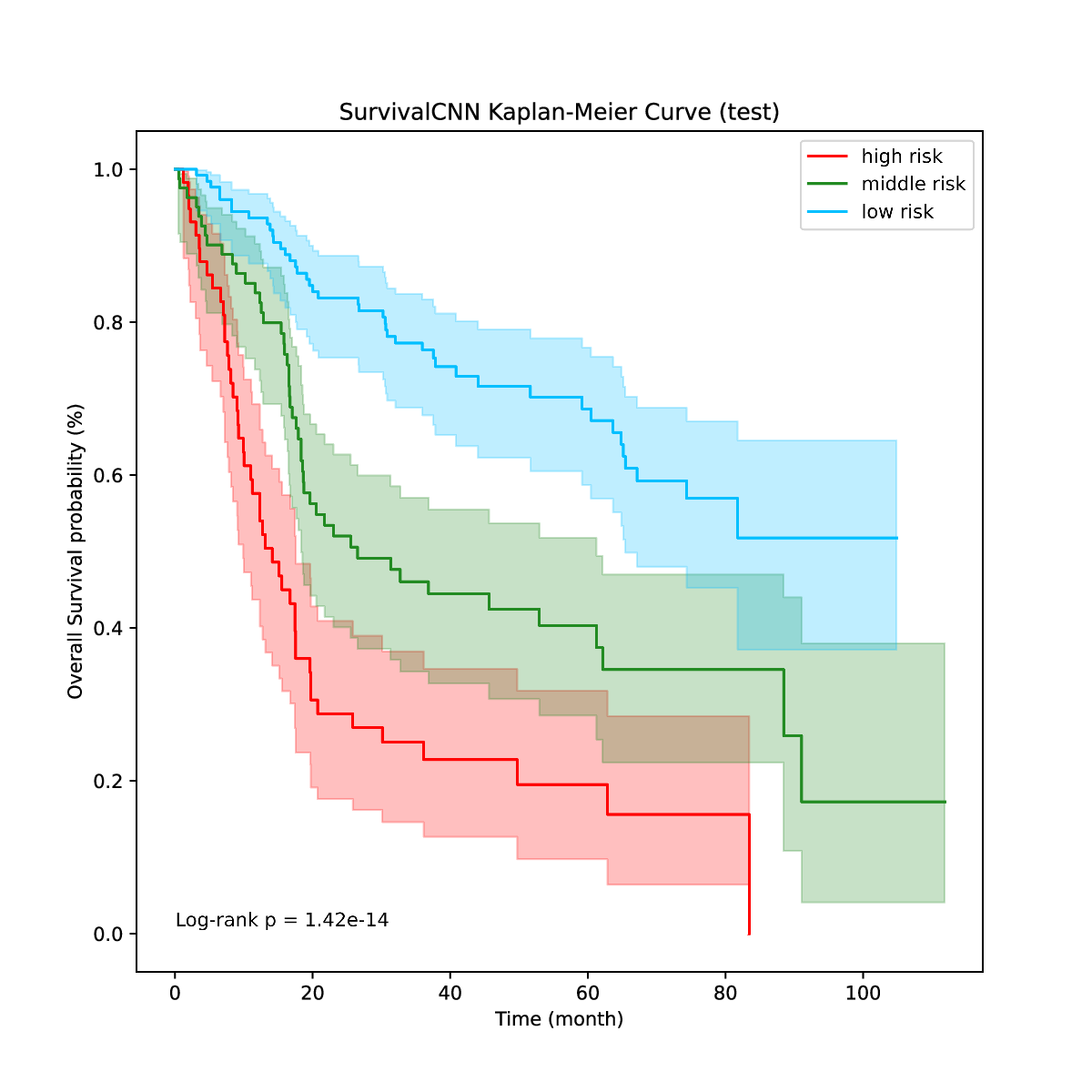}}
\quad
\subfloat[CACA-UCOM]{\includegraphics[width=0.165\textwidth]{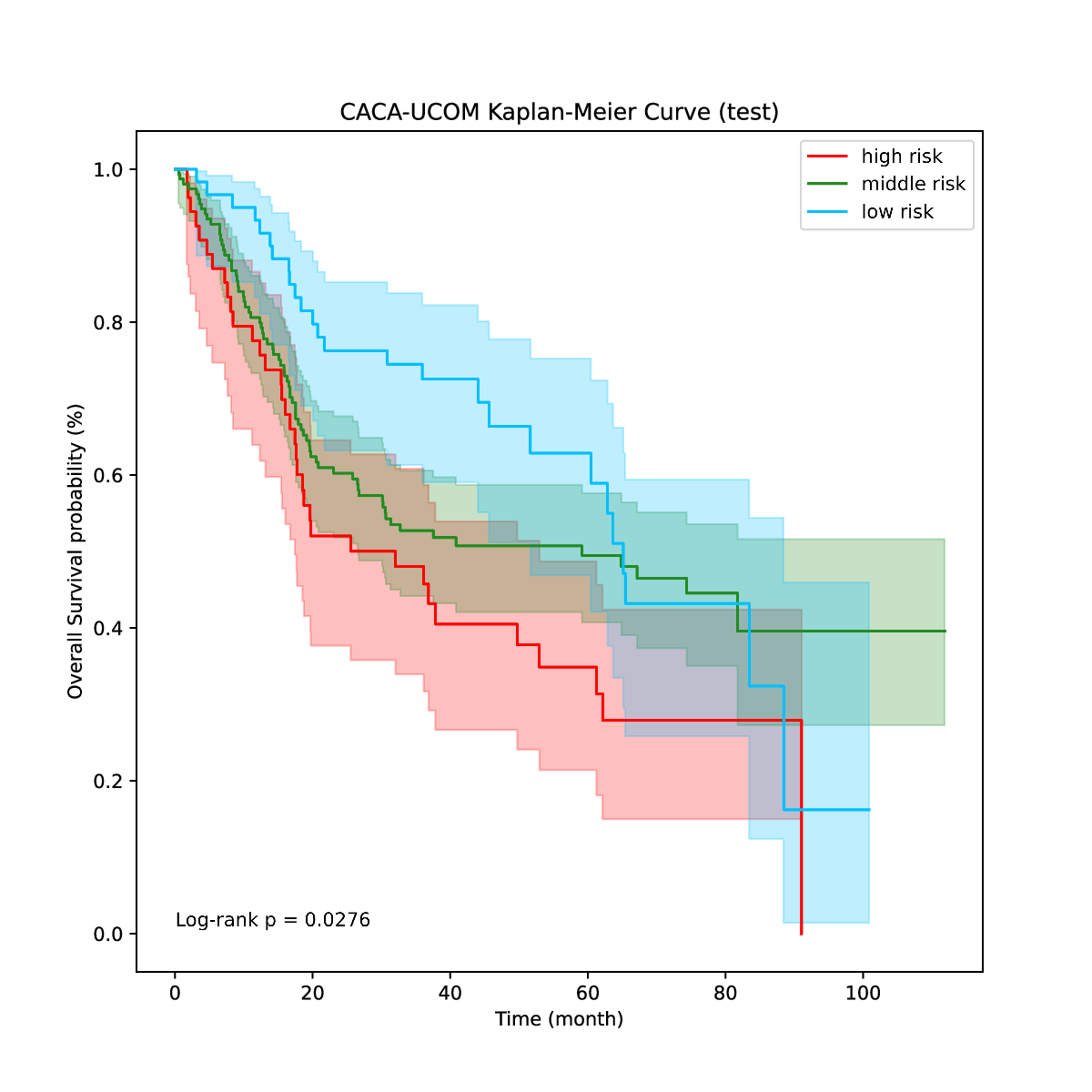}}
\subfloat[FullerMDA]{\includegraphics[width=0.165\textwidth]{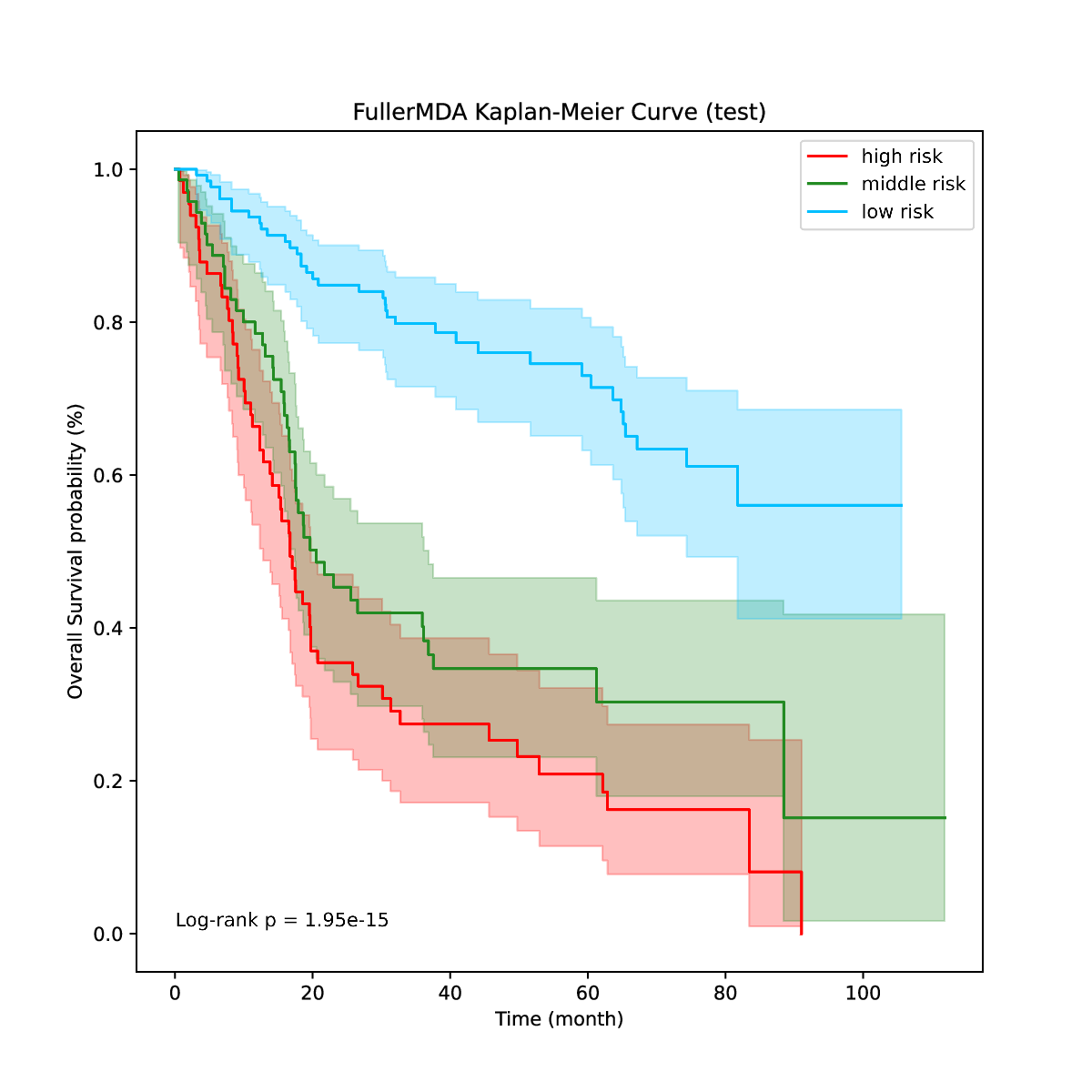}}
\subfloat[MIFI-AE]{\includegraphics[width=0.165\textwidth]{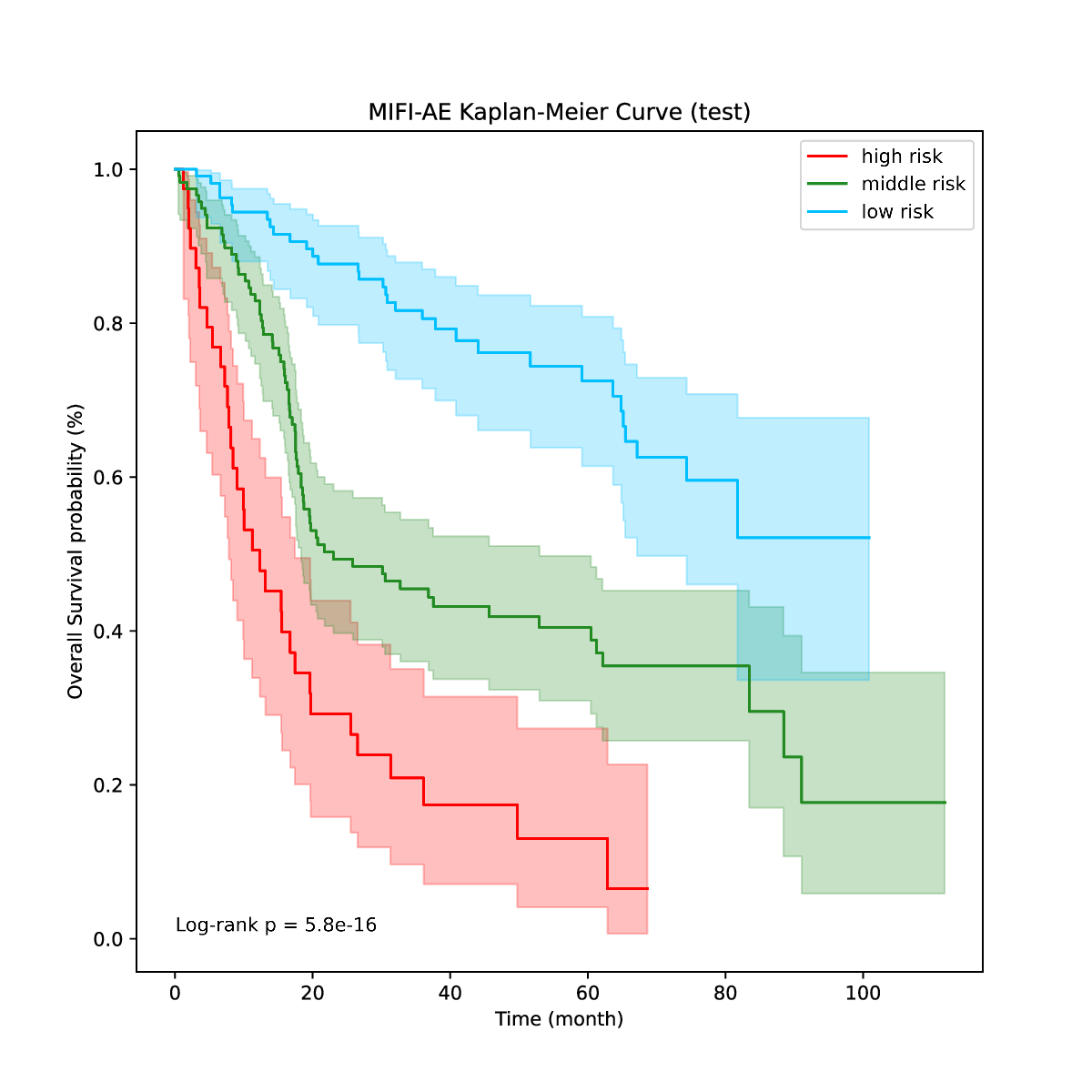}}
\caption{Kaplan-Meier (KM) curves of OS for compared and proposed models.}
\label{km_os}
\end{figure}

For the purpose of further validating the clinical applicability of the models, the Kaplan-Meier (KM) method was applied by stratifying ESCC patients into three risk subgroups using the cut-off threshold automatically analyzed by X-tile software \cite{bland1998survival,camp2004x}. The log-rank test was utilized to detect whether the distinction between curves is statistically significant. Here, we pick one of the test folds for illustration purposes. In Fig. \ref{km_os}, our MIFI-AE achieved the lowest log-rank test p-value (p = 5.8e-16), which demonstrates that our model has the best risk group stratification and clinical decision-making support ability.

\section{Ablation Study}
\subsection{Effectiveness of Proposed Modules}
To verify the effectiveness of the proposed sub-modules, the CMIFM and MFFSM were used to test the C-index performance. The details of the ablated models can be seen in Table \ref{tab:ablated_models}. It can be seen from the fact that models that included only uni-modules, such as MIFI-AE without CMIFM (0.692 $\pm$ 0.02) and MIFI-AE without MFFSM (0.686 $\pm$ 0.01), get a higher C-index performance than MIFI-AE without both (0.682 $\pm$ 0.01), indicating that both CMIFM and MFFSM contribute to the final performance. Our MIFI-AE, which included both CMIFM and MFFSM, reached the best C-index performance (0.697 $\pm$ 0.02), indicating that the combination mechanism of both modules can function effectively.

\begin{table}[!htbp]
    \centering
    \caption{C-index performance for ablation studies.}
     \begin{tabular}{cccc}
  \hline\rule{0pt}{5pt}
  \multirow{2}*{Methods}  & \multicolumn{2}{c}{Module included} & \multirow{2}*{\makecell[c]{C-index \\ (Mean $\pm$ Std)}} \\ 
  \cline{2-3}\rule{0pt}{8pt}
   & CMIFM & MFFSM   \\
  \hline\rule{0pt}{10pt}
  \makecell[c]{MIFI-AE \\ (w/o CMIFM)}   & $\checkmark$ &  & 0.692 $\pm$ 0.02  \\
  \makecell[c]{MIFI-AE \\ (w/o MFFSM)} &  & $\checkmark$ & 0.686 $\pm$ 0.01 \\
  \makecell[c]{MIFI-AE \\(w/o Both)} &  &  &  0.682 $\pm$ 0.01 \\
  \hline\rule{0pt}{10pt}
  \textbf{\makecell[c]{MIFI-AE  (Ours)}} & $\checkmark$ & $\checkmark$ & \textbf{0.697 $\pm$ 0.02 }  \\ 
  \hline
    \end{tabular}
    \label{tab:ablated_models}
\end{table}

\subsection{Effectiveness of Proposed Alignment Loss}
To validate the capability of the proposed MJ-Loss, we conduct the ablation experiment for alignment loss $\mathcal{L}_{\mathrm{Align}}$. The results can be seen in Table \ref{tab:ablated_loss}. As can be seen, the proposed model acquired the best C-index under optimization of MJ-Loss (w/ $\mathcal{L}_{\mathrm{Align}}$); the loss without alignment loss (w/o $\mathcal{L}_{\mathrm{Align}}$) performs slightly worse than MJ-Loss, which indicates that the alignment loss is indeed helpful for eliminating the semantic gap between multi-modal representations and multi-modal feature learning.

\begin{table}[!htbp]
    \centering
    \caption{C-index performance for ablation studies.}
     \begin{tabular}{cccc}
  \hline\rule{0pt}{5pt}
  \multirow{2}*{Methods}  & \multicolumn{2}{c}{Loss included} & \multirow{2}*{\makecell[c]{C-index \\ (Mean $\pm$ Std)}} \\ 
  \cline{2-3}\rule{0pt}{8pt}
   & w/ $\mathcal{L}_{\mathrm{Align}}$ & w/o $\mathcal{L}_{\mathrm{Align}}$   \\
  \hline\rule{0pt}{10pt}
      \makecell[c]{MIFI-AE}  &  & $\checkmark$ &  0.691 $\pm$ 0.02 \\
  \textbf{\makecell[c]{MIFI-AE  (Ours)}} & $\checkmark$ &  & \textbf{0.697 $\pm$ 0.02 }  \\ 
  \hline
    \end{tabular}
    \label{tab:ablated_loss}
\end{table}
\section{Conclusion}
In this paper, we innovatively propose an autoencoder-based survival prediction model to predict the OS of ESCC patients. Different from previous models, we propose two novel modules for cross-modal feature reinforcement and multi-scale feature map fusion. By applying these two modules, our model can extract prognosis-related features more effectively and has a stronger ability to extract high-dimensional latent features from CT images. Furthermore, an MJ-loss was proposed to eliminate the cross-modal semantic gap and optimize the model. The experiment results show that our model performs best at discriminative ability and risk stratification, which indicates that our model can be utilized in clinical decision-making.

\section{Compliance with Ethical Standards}
This study was performed in line with the principles of the Declaration of Helsinki. Approval was granted by the Institutional Ethics Committee of Sichuan Cancer Hospital (No. SCCHEC-02-2020-015).

\section{Acknowledgments}
This work was supported by National Natural Science Foundation of China under Grant No. 62201323, Natural Science Foundation of Jiangsu Province under Grant No. BK20220266, Zhejiang Provincial Natural Science Foundation of China under Grant No. LDT23F01015F01 and Science and Technology Department of Sichuan Province (2023YFS0488 and 2023YFQ0055).
\bibliographystyle{IEEEbib}
\bibliography{refs}

\end{document}